\documentclass{moriond}

\def\citet{\cite}

\def\be{\begin{equation}}
\def\ee{\end{equation}}
\def\bea{\begin{eqnarray}}
\def\eea{\end{eqnarray}}

\begin{document}
\vspace*{4cm}

\title{Flow in pPb collisions at 5 TeV?}

\author{K.$\,$Werner$^{(a)}$, B. Guiot$^{(a)}$, Iu.$\,$Karpenko$^{(b,c)}$,
T.$\,$Pierog$^{(d)}$\vspace{0.3cm}}

\address{$^{(a)}$ SUBATECH, University of Nantes -- IN2P3/CNRS-- EMN, Nantes,
France}

\address{$^{(b)}$ Bogolyubov Institute for Theoretical Physics, Kiev 143,
03680, Ukraine}

\address{$^{(c)}$ FIAS, Johann Wolfgang Goethe Universitaet, Frankfurt am
Main, Germany}

\address{$^{(d)}$Karlsruhe Inst. of Technology, KIT, Campus North, Inst.
f. Kernphysik, Germany}

\maketitle\abstracts{
There is little doubt that hydrodynamic flow has been observed in
heavy ion collisions at the LHC and RHIC, mainly based on results
on azimuthal anisotropies, but also on particle spectra of identified
particles, perfectly compatible with hydrodynamic expansions. Surprisingly,
in p-Pb collisions one observes a very similar behavior. So do we
see flow even in p-Pb? We will try to answer this question.
}

Collective hydrodynamic flow seems to be well established in heavy
ion (HI) collisions at energies between 200 and 2760 AGeV, whereas
p-p and p-nucleus (p-A) collisions are often considered to be simple
reference systems, showing {}``normal'' behavior, such that deviations
of HI results with respect to p-p or p-A reveal {}``new physics''.
Surprisingly, the first results from p-Pb at 5 TeV on the transverse
momentum dependence of azimuthal anisotropies and particle yields
are very similar to the observations in HI scattering \citet{cms,alice}. 

Do we see radial flow in p-Pb collisions? In order to answer this
question, we will employ the EPOS3 approach \citet{epos3}, well suited
for this problem, since it provides within a unique theoretical scheme
the initial conditions for a hydrodynamical evolution in p-p, p-A,
and HI collisions. The initial conditions are generated in the Gribov-Regge
multiple scattering framework. An individual scattering is referred
to as Pomeron, identified with a parton ladder, eventually showing
up as flux tubes (also called strings). Each parton ladder is composed
of a pQCD hard process, plus initial and final state linear parton
emission. Our formalism is referred to as {}``Parton based Gribov
Regge Theory'' and described in very detail in \citet{hajo}. Based
on these initial conditions, we performed already ideal hydrodynamical
calculations (EPOS2) \citet{epos2,jetbulk,kw1,kw2} to analyse HI
and p-p scattering at RHIC and LHC. In EPOS3 we add two major improvements:
a more sophisticated treatment of nonlinear effects in the parton
evolution by considering individual (per Pomeron) saturation scales
\citet{sat1}, and a 3D+1 viscous hydrodynamical evolution. There
are also changes in our core-corona procedure, which %
\begin{figure}[tb]
\begin{centering}
\includegraphics[angle=270,scale=0.24]{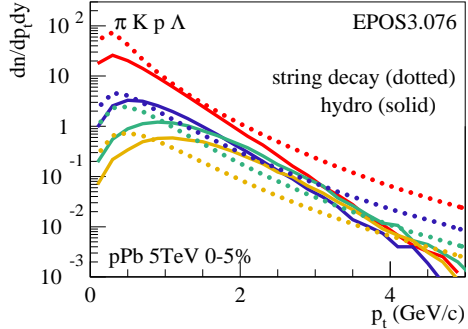}
\par\end{centering}

\caption{(Color online) Identified particle spectra as a function of $p_{t}$,
for central (0-5\%) p-Pb collisions at 5.02 TeV. We show results for
particle production from string decay, i.e. EPOS without hydro (dotted
curves), and particle production from pure hydro, without corona (solid
lines). In both cases, we show (from top to bottom) pions, kaons,
protons, and lambdas.\label{fig:flow}}
\end{figure}
amounts to separate the initial energy of the flux tubes into a part
which constitutes the initial conditions for hydro (core) and the
particles which leave the {}``matter''. This is crucial as well
in proton-nucleus collisions (as in all other collision types).

To understand the results discussed later in this paper, we show in
fig. \ref{fig:flow} the effect of flow on identified particle spectra,
by comparing $p_{t}$ distributions from pure string decay to spectra
from a pure hydrodynamic evolution. In case of string fragmentation,
heavier particles are strongly suppressed compared to lighter ones,
but the shapes are not so different. This picture changes completely
in the fluid case: The heavier the particle, the more it gets shifted
to higher $p_{t}$. This is a direct consequence of the fact that
the particles are produced from fluid cells characterized by radial
flow velocities, which gives more transverse momentum to heavier particles.

There are few other studies of hydrodynamic expansion in proton-nucleus
systems. In \citet{hydro-bozek}, fluctuating initial conditions based
on the so-called Monte Carlo Glauber model (which is actually a wounded
nucleon model) are employed, followed by a viscous hydrodynamical
evolution. Also \citet{hydro-schenke} uses fluctuating initial conditions,
here based on both Glauber Monte Carlo and Glasma initial conditions.
Finally in \citet{hydro-qin}, ideal hydrodynamical calculations are
performed, starting from smooth Glauber model initial conditions. 

In the following, we will compare experimental data on identified
particle production with our simulation results (referred to as EPOS3),
and in addition to some other models, as there are QGSJETII \citet{qgsjet},
AMPT \citet{ampt}, and EPOS$\,$LHC \citet{eposlhc}. The QGSJETII
model is also based on Gribov-Regge multiple scattering, but there
is no fluid component. The main ingredients of the AMPT model are
a partonic cascade and then a hadronic cascade, providing in this
way some {}``collectivity''. EPOS$\,$LHC is a tune (using LHC data)
of EPOS1.99. As all EPOS1 models, it contains flow, put in by hand,
parametrizing the collective flow at freeze-out. Finally, the approach
discussed in this paper (EPOS3) contains a full viscous hydrodynamical
simulation. So it is interesting to compare these four models, since
they differ considerably concerning the implementation of flow, from
full hydrodynamical flow in EPOS3 to no flow in QGSJETII.

\begin{figure}[tb]
\begin{minipage}[t]{0.45\columnwidth}%
\begin{minipage}[c]{1\columnwidth}%
\vspace*{-0.2cm}

\hspace*{-0.cm}\includegraphics[angle=270,scale=0.18]{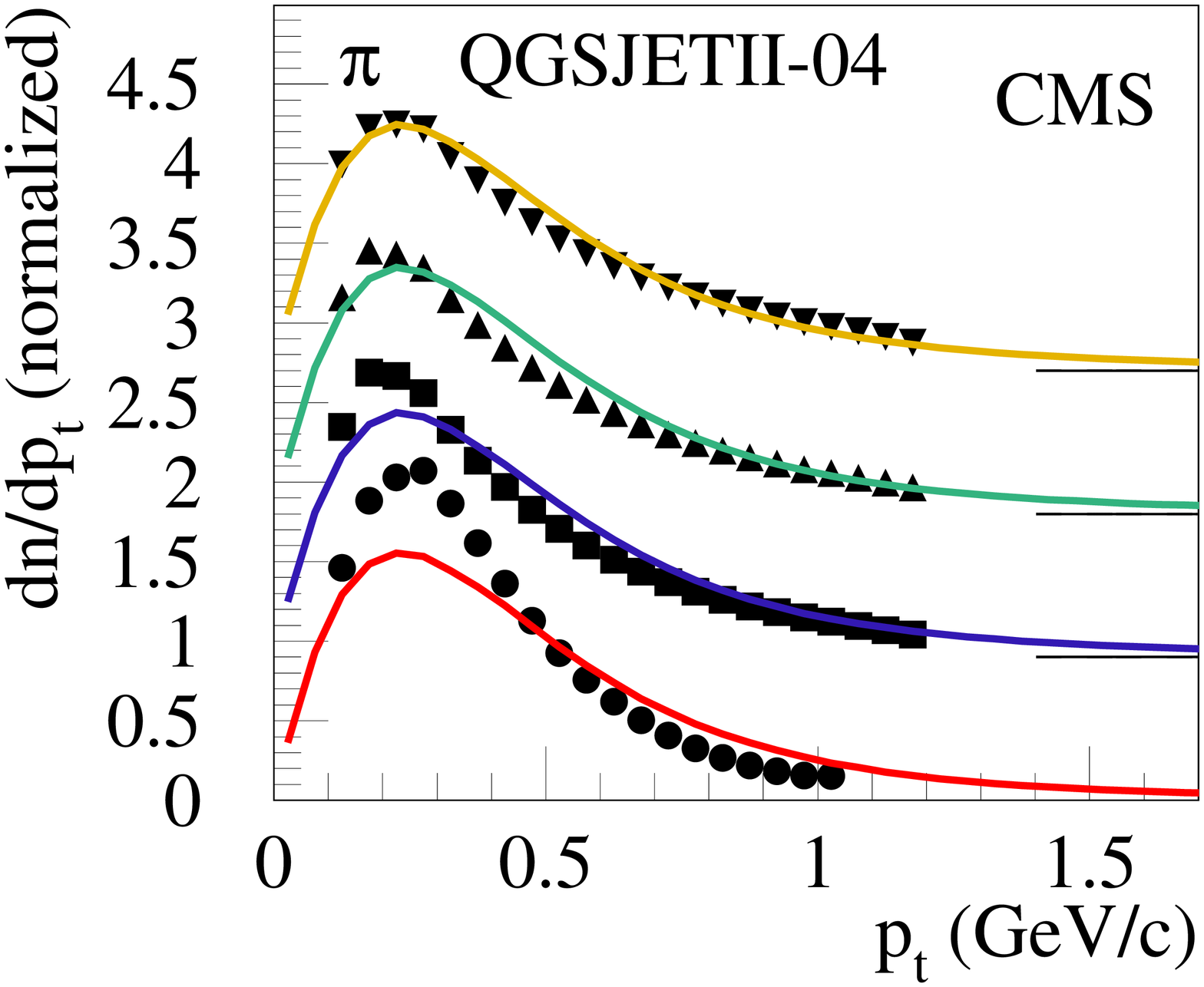}\hspace*{-0.5cm}\includegraphics[angle=270,scale=0.18]{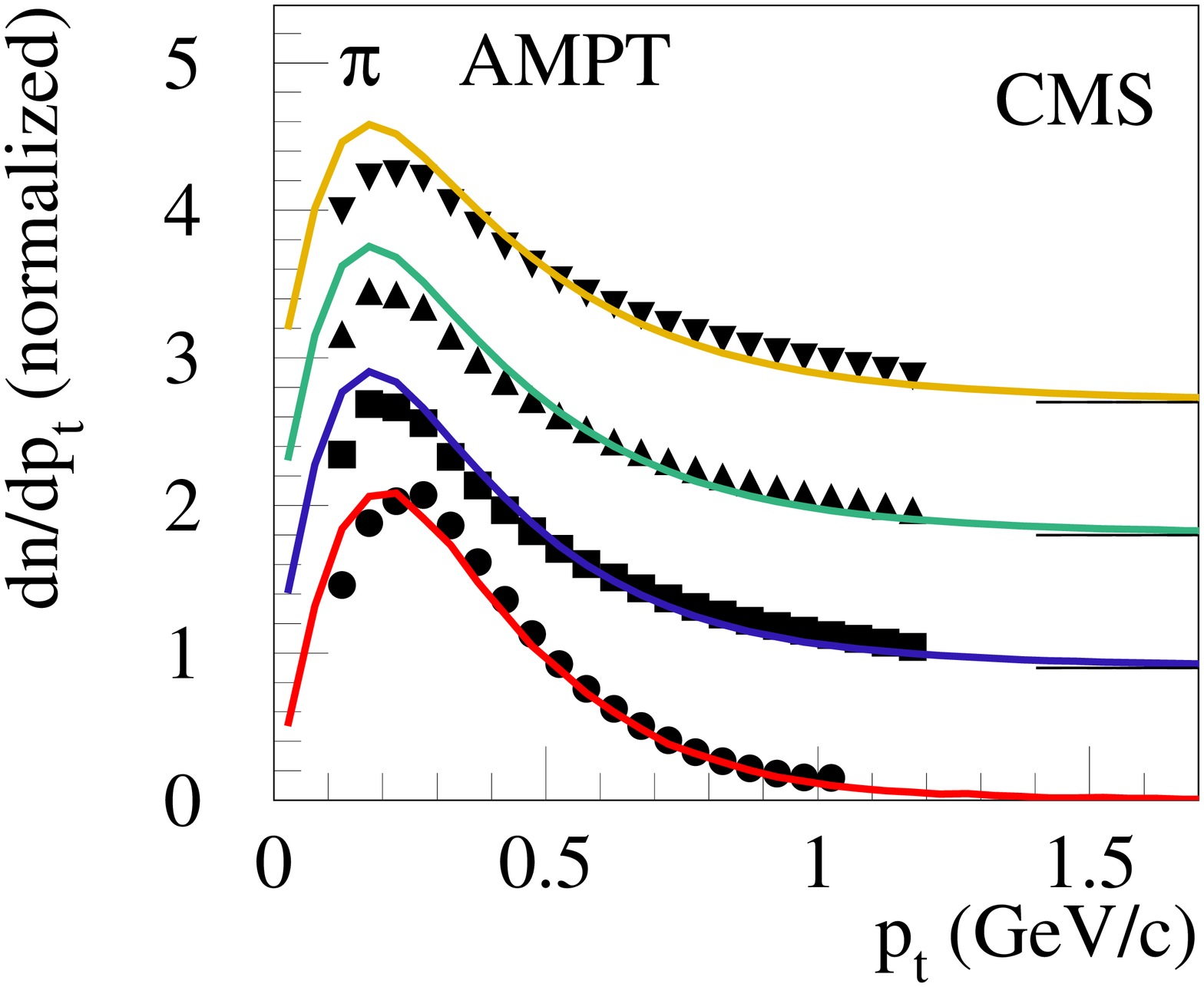}\vspace*{-0.2cm}

\hspace*{-0.cm}\includegraphics[angle=270,scale=0.18]{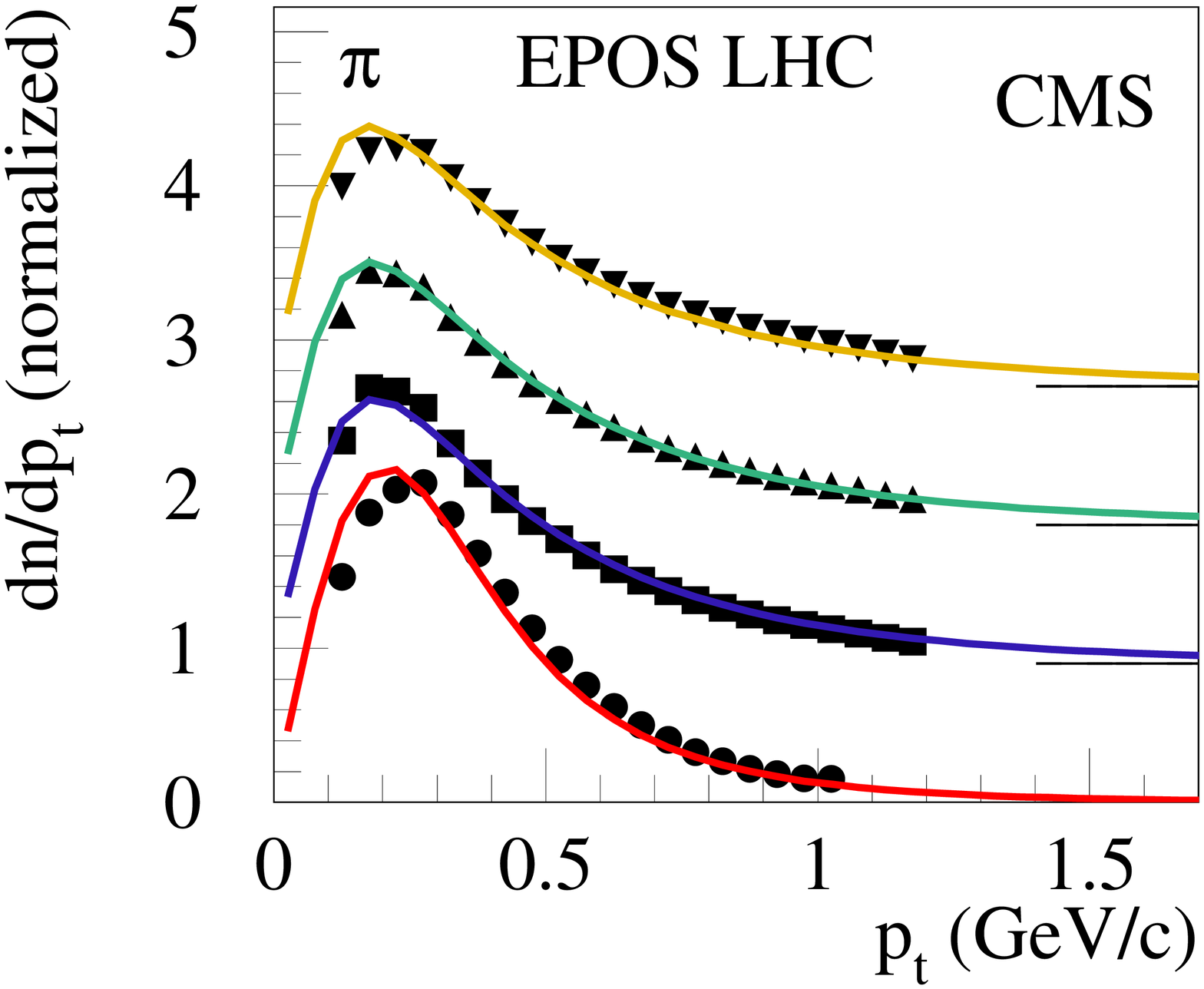}\hspace*{-0.5cm}\includegraphics[angle=270,scale=0.18]{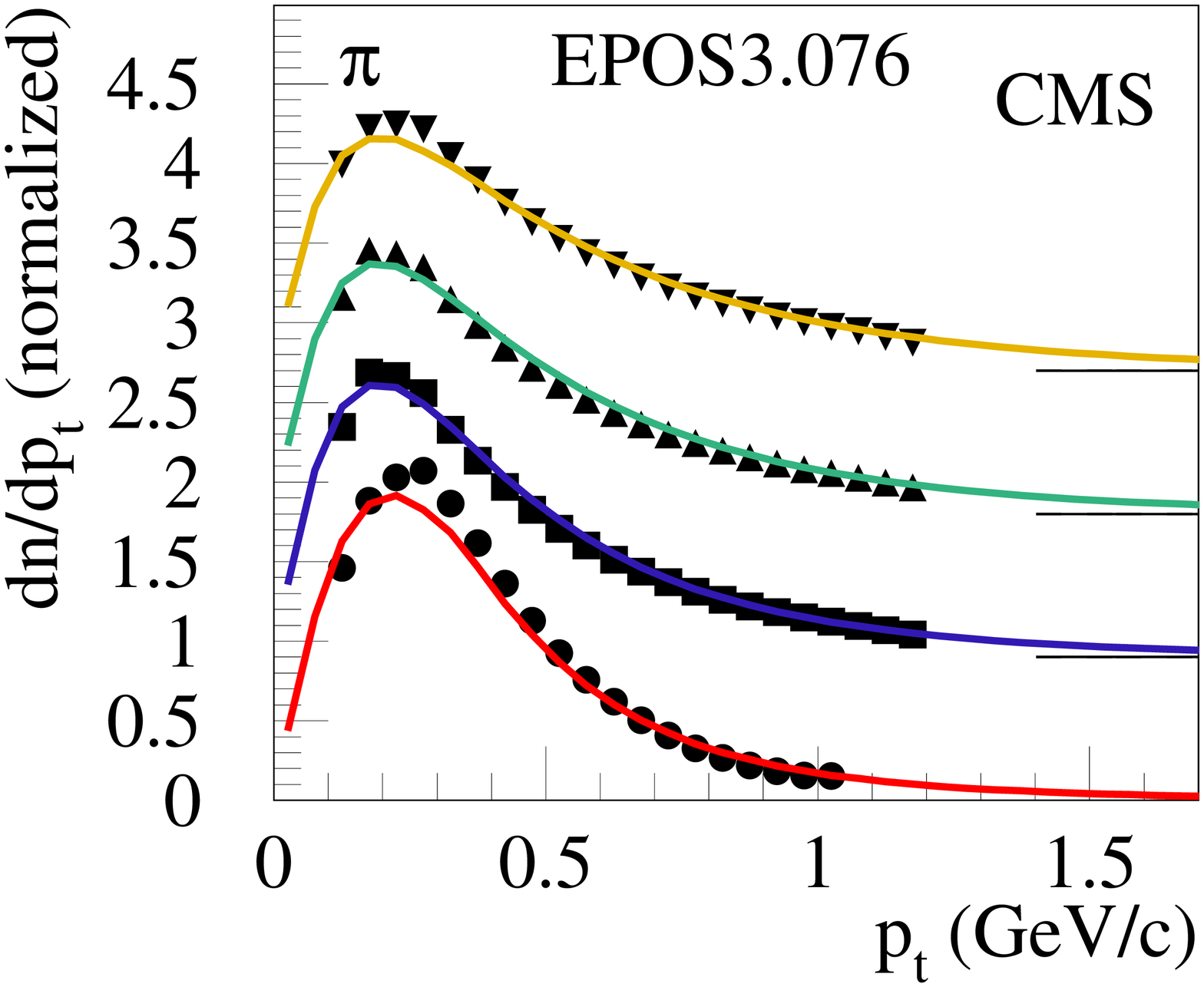}%
\end{minipage}

\noindent \caption{(Color online) Transverse momentum spectra of pions in p-Pb scattering
at 5.02 TeV, for four different multiplicity classes with mean values
(from bottom to top) of 8, 84, 160, and 235 charged tracks. \label{fig:cms1}}
\end{minipage}$\qquad\quad$%
\begin{minipage}[t]{0.45\columnwidth}%
\begin{minipage}[c]{1\columnwidth}%
\vspace*{-0.2cm}

\hspace*{-0.3cm}\includegraphics[angle=270,scale=0.18]{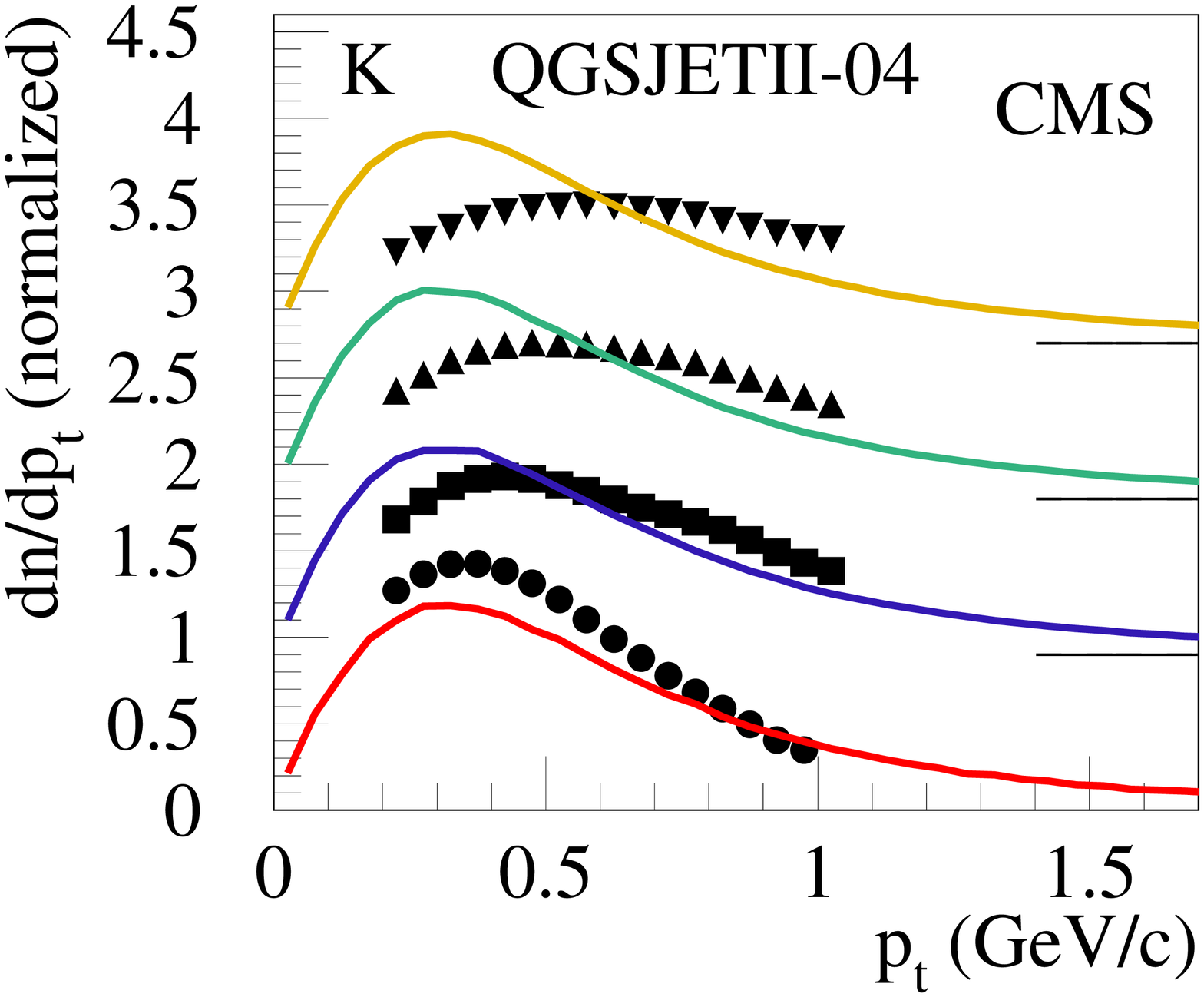}\hspace*{-0.5cm}\includegraphics[angle=270,scale=0.18]{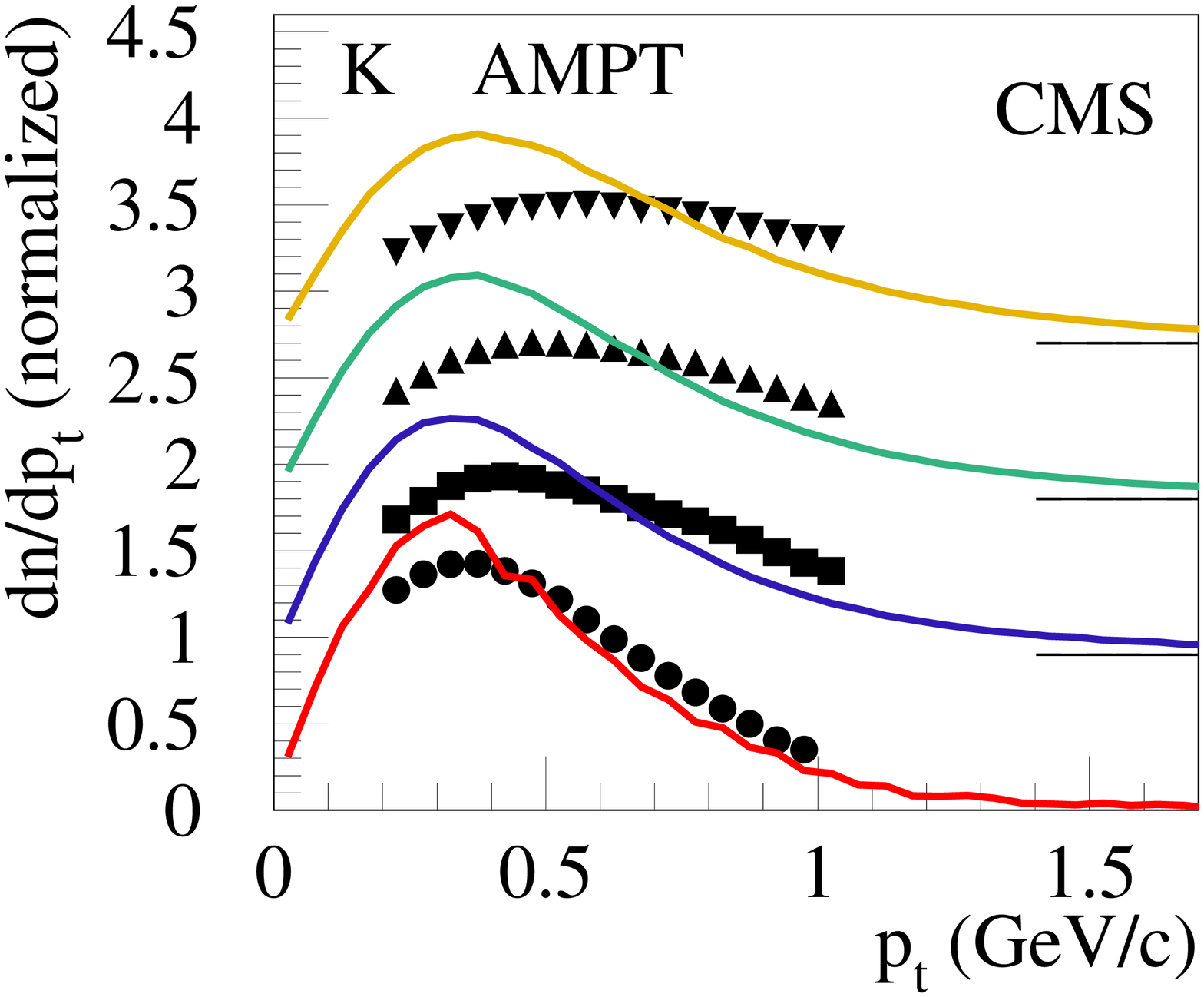}\vspace*{-0.2cm}

\hspace*{-0.3cm}\includegraphics[angle=270,scale=0.18]{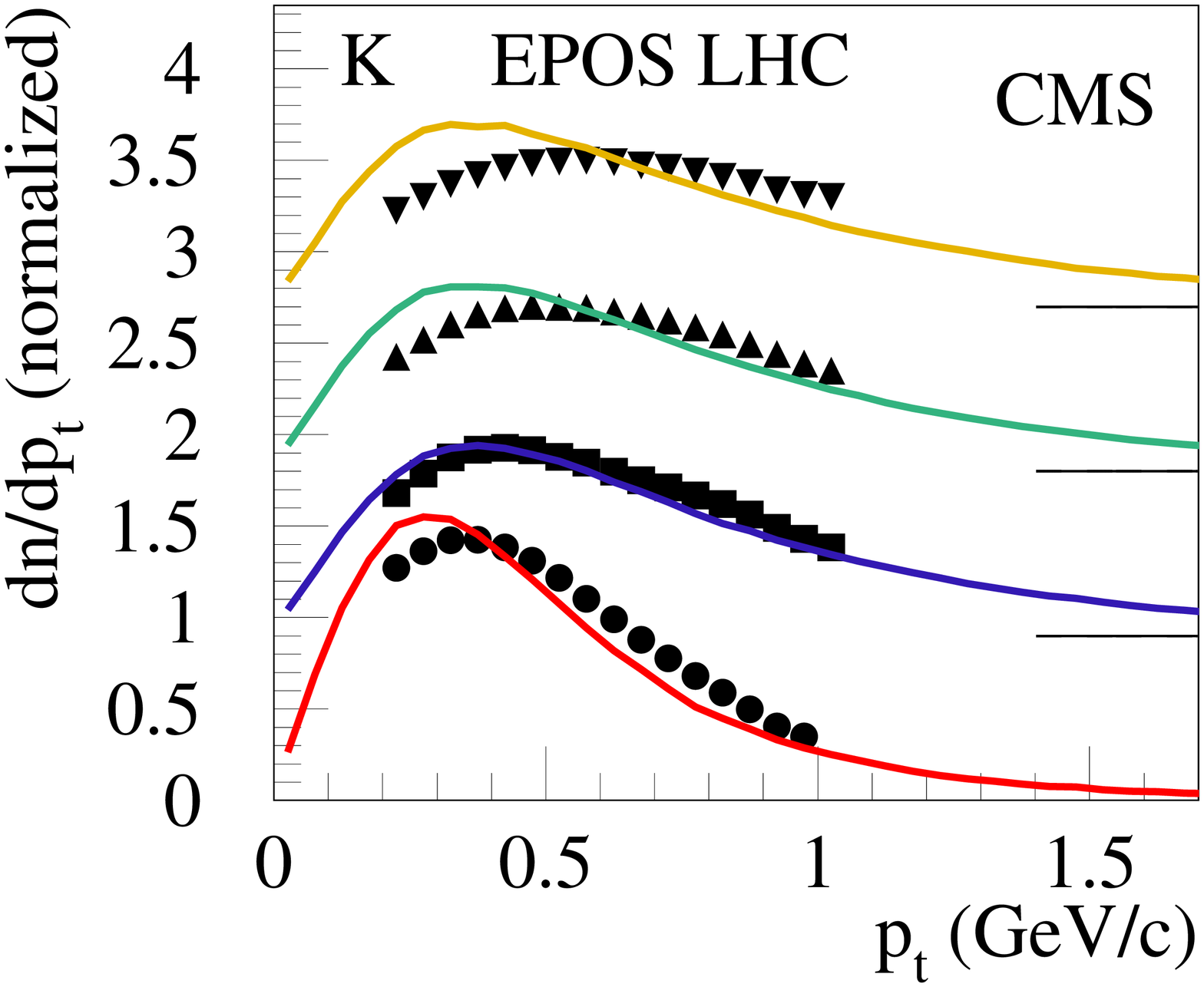}\hspace*{-0.5cm}\includegraphics[angle=270,scale=0.18]{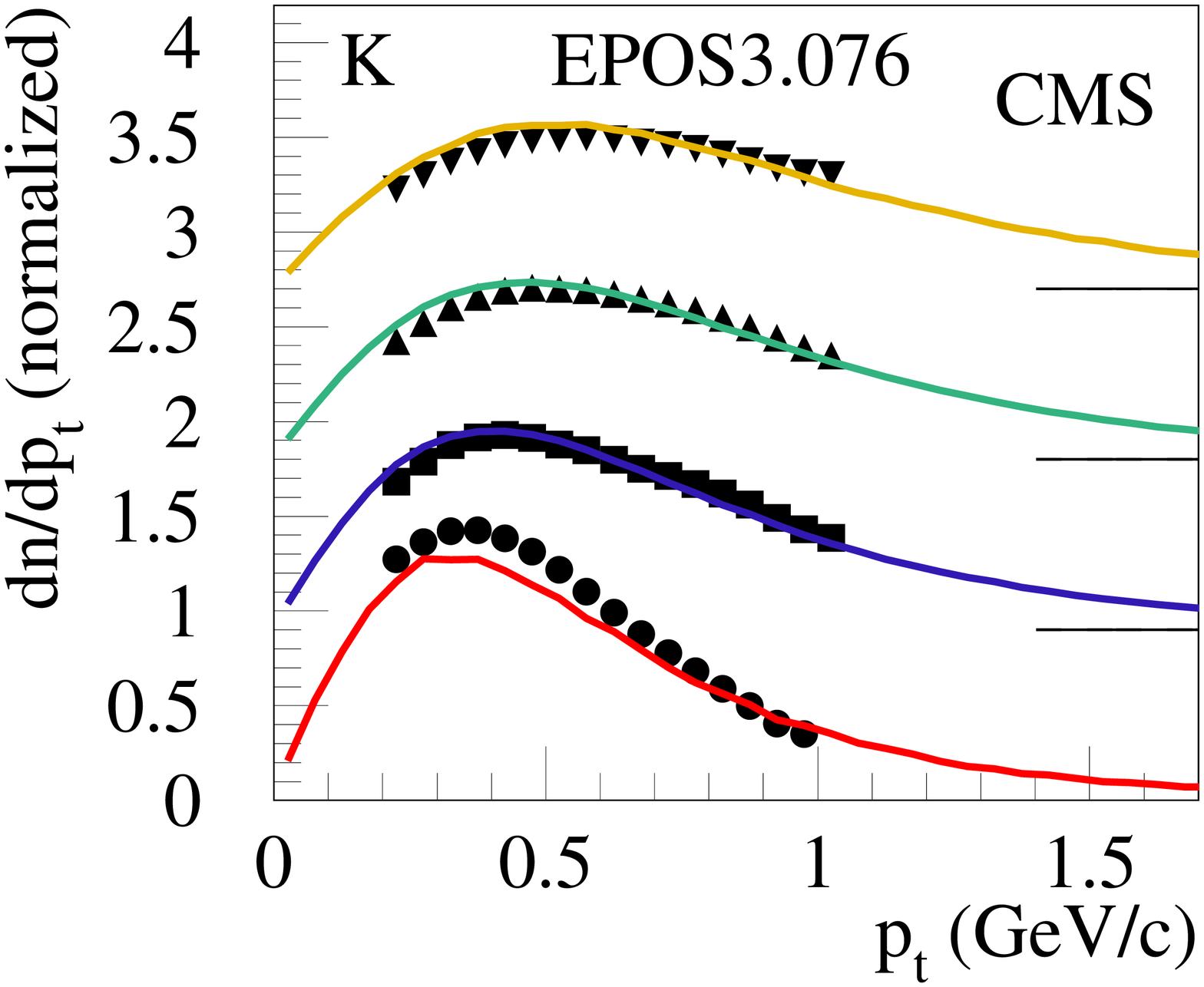}%
\end{minipage}

\noindent \caption{(Color online) Same as fig. \ref{fig:cms1}, but for kaons. We show
data from CMS  (symbols) and simulations from QGSJETII,
AMPT, EPOS$\,$LHC, and EPOS3, as indicated in the figures. \label{fig:cms2} }
\end{minipage}\vspace{0.5cm}

\begin{minipage}[c]{0.4\columnwidth}%
\begin{minipage}[c]{1\columnwidth}%
\vspace*{-0.2cm}

\hspace*{-0.cm}\includegraphics[angle=270,scale=0.18]{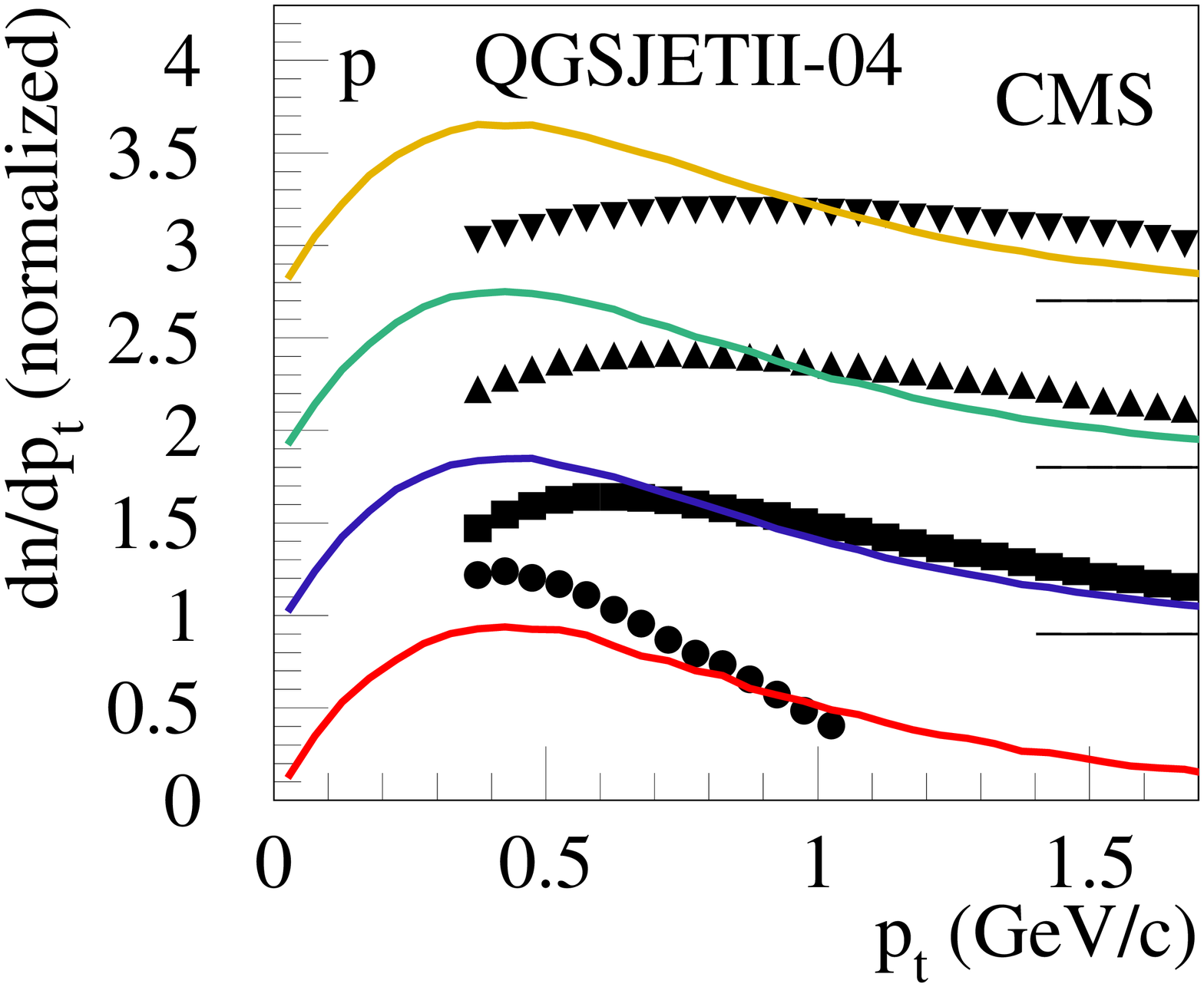}\hspace*{-0.5cm}\includegraphics[angle=270,scale=0.18]{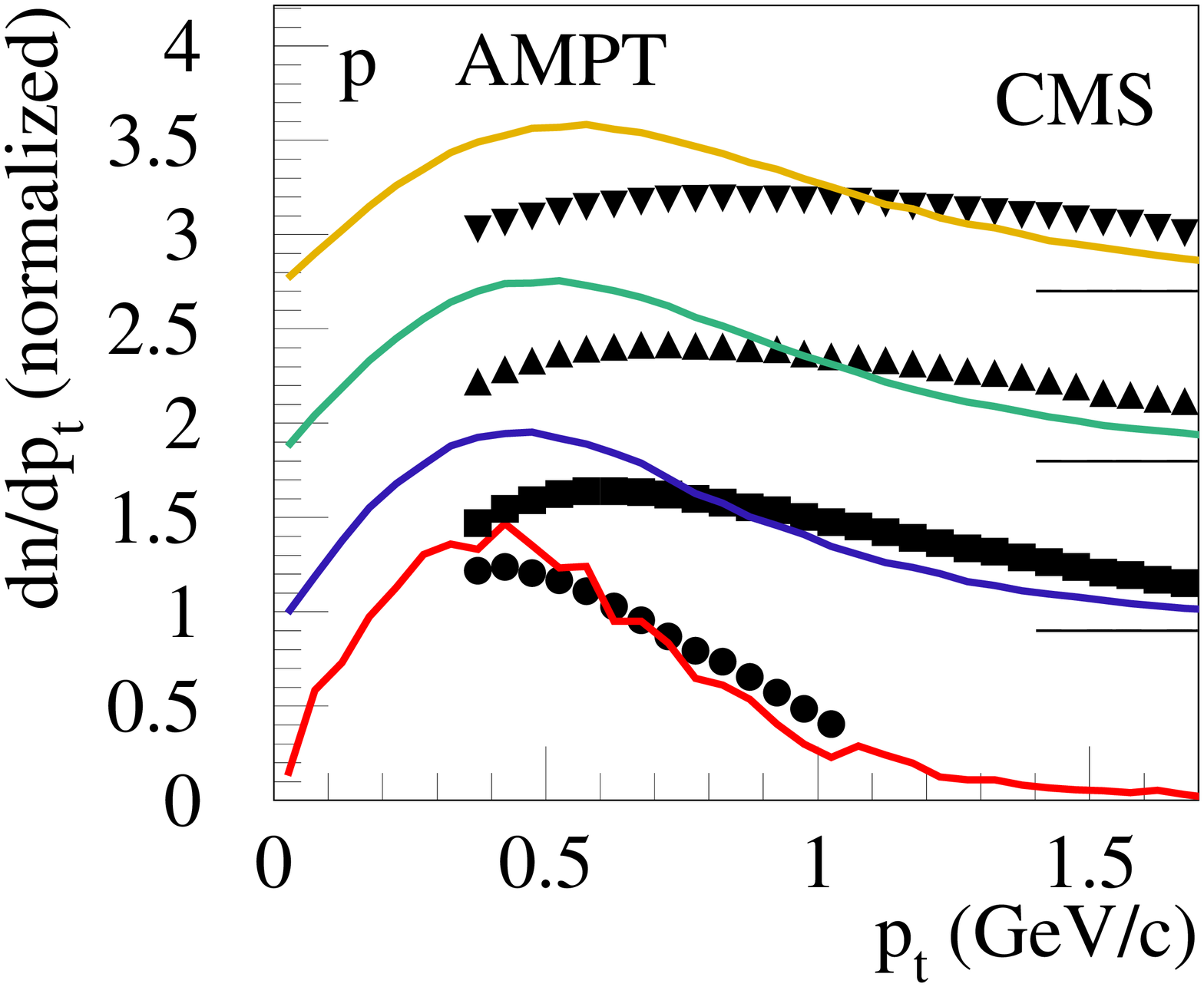}\vspace*{-0.2cm}

\hspace*{-0.cm}\includegraphics[angle=270,scale=0.18]{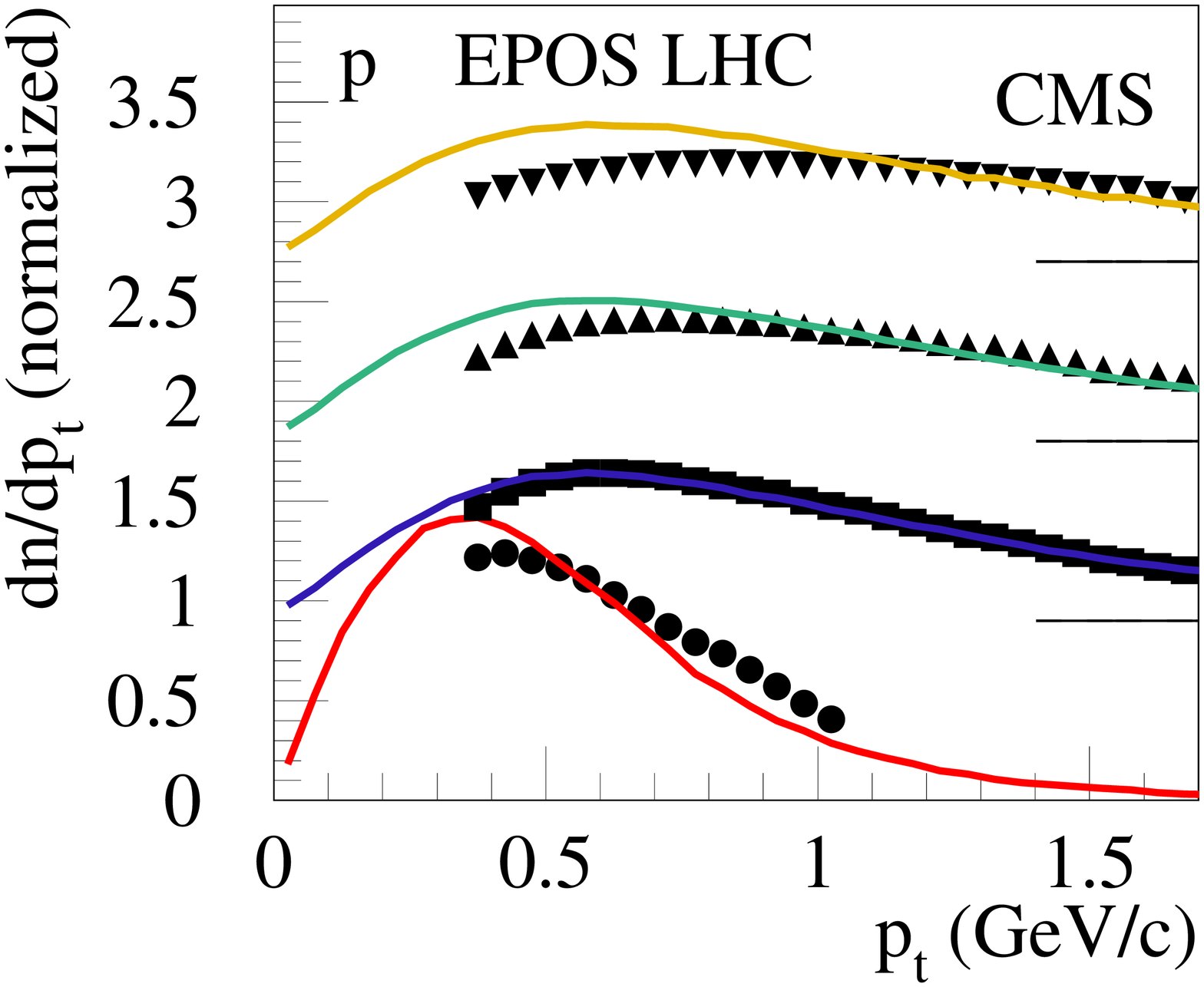}\hspace*{-0.5cm}\includegraphics[angle=270,scale=0.18]{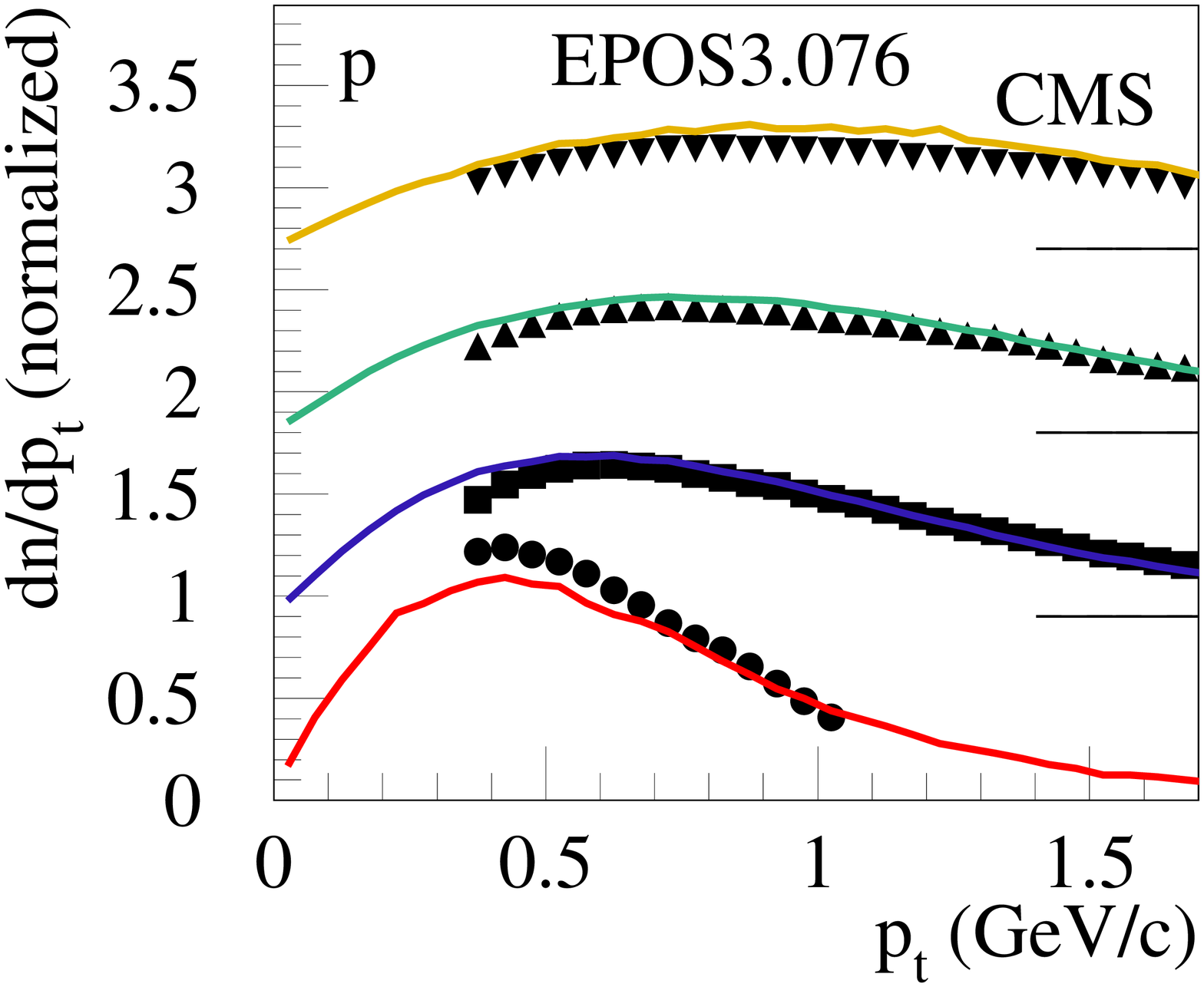}%
\end{minipage}

\noindent \caption{(Color online) Same as fig. \ref{fig:cms1}, but for protons.\label{fig:cms3}\protect \\
$\,$\protect \\
$\,$\protect \\
$\:$\protect \\
$\;$}
\end{minipage}$\qquad\quad$%
\begin{minipage}[c]{0.55\columnwidth}%
\begin{flushright}
\begin{minipage}[c]{1\columnwidth}%
\vspace*{-0.2cm}

\hspace*{0.3cm}\includegraphics[angle=270,scale=0.18]{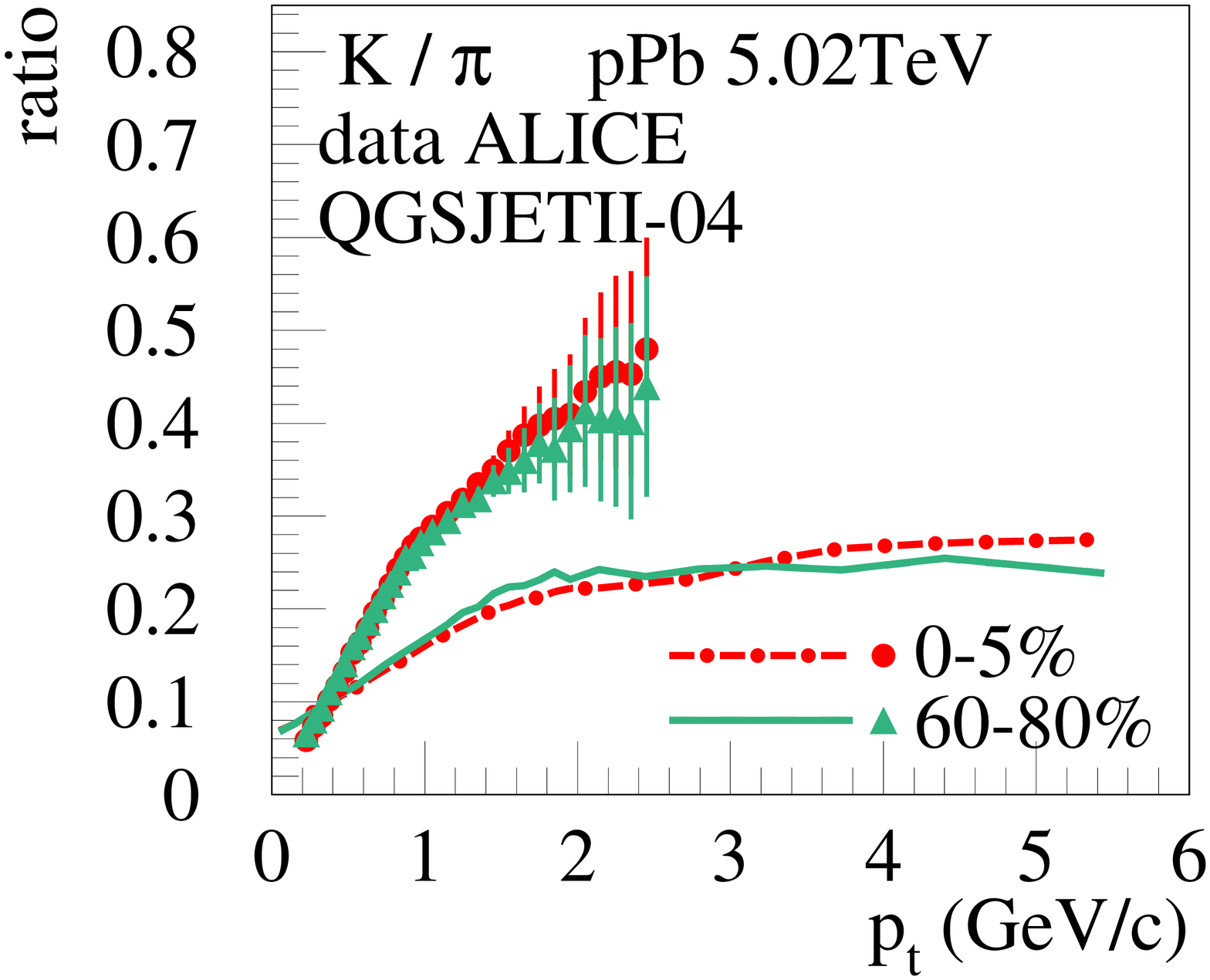}\hspace*{-0.5cm}\includegraphics[angle=270,scale=0.18]{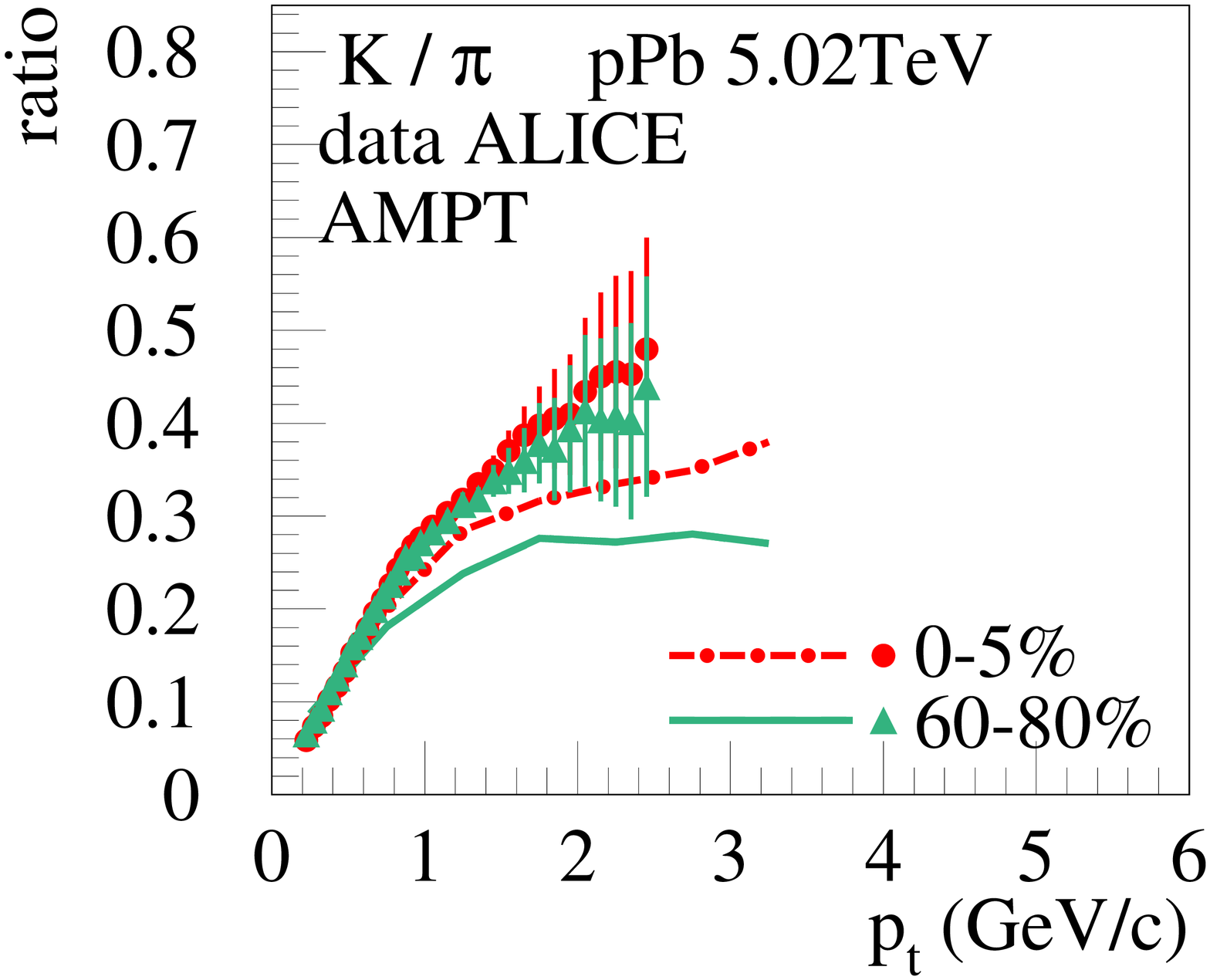}\vspace*{-0.2cm}

\hspace*{0.3cm}\includegraphics[angle=270,scale=0.18]{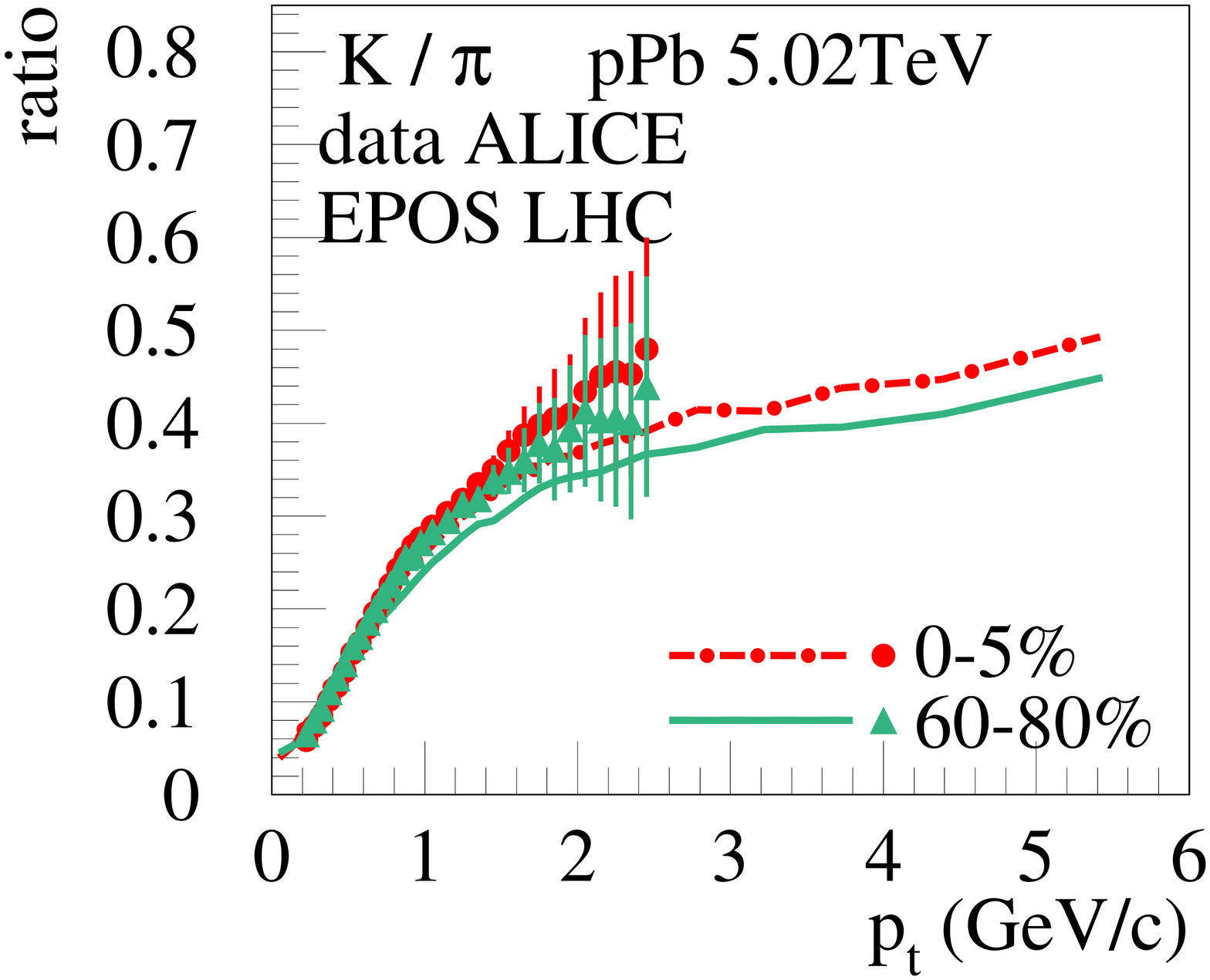}\hspace*{-0.5cm}\includegraphics[angle=270,scale=0.18]{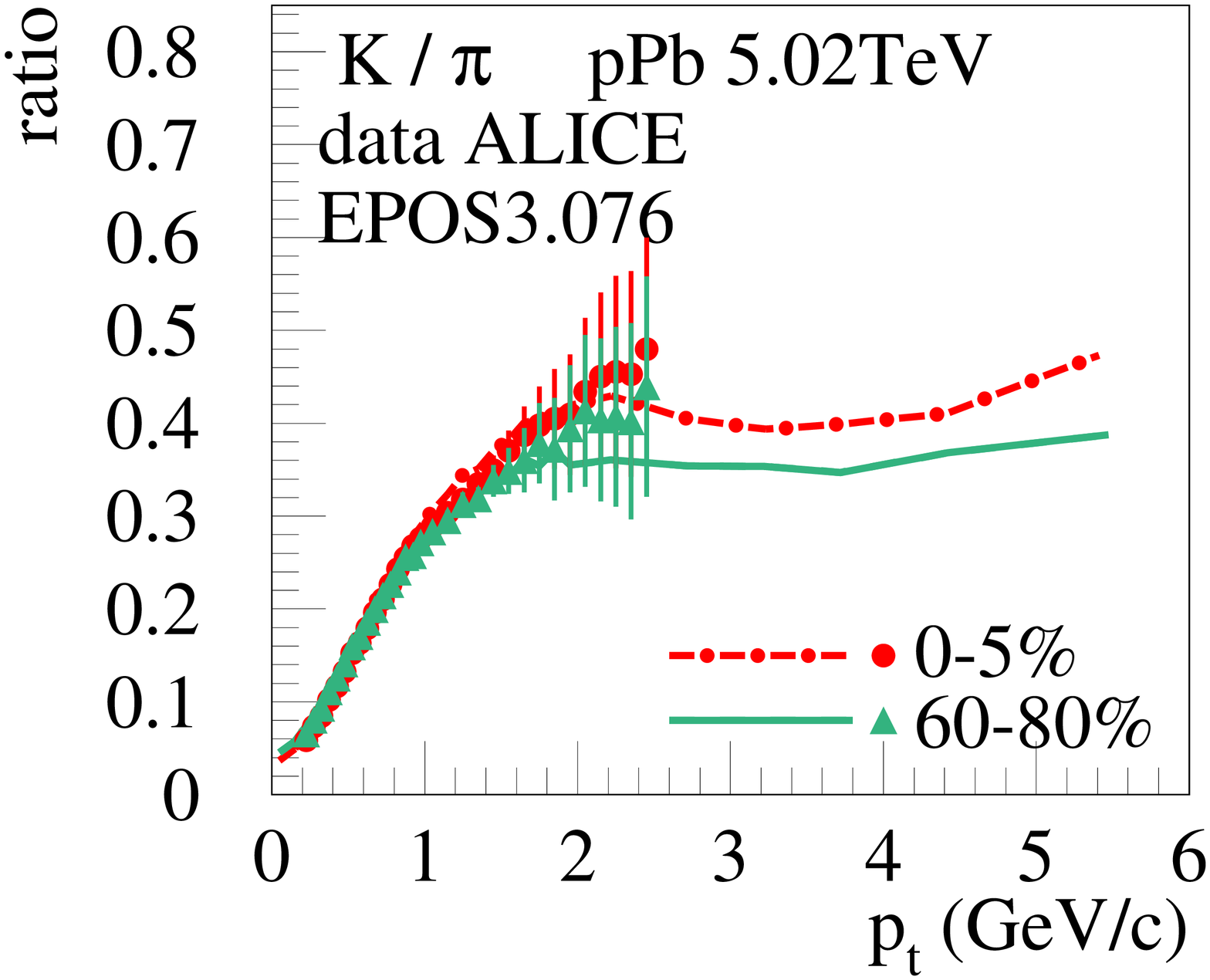}%
\end{minipage}
\par\end{flushright}

\noindent \caption{(Color online) Kaon over pion ratio as a function of transverse momentum
in p-Pb scattering at 5.02 TeV, for the 0-5\% highest multiplicity
(red dashed-dotted lines, circles) and 60-80\% (green solid lines,
triangles).\label{fig:selid07} \protect \\
~}
\end{minipage}%
\end{figure}
The CMS collaboration published a detailed study \citet{cms} of the
multiplicity dependence of (normalized) transverse momentum spectra
in p-Pb scattering at 5.02 TeV. The multiplicity (referred to as $N_{\mathrm{track}}$)
counts the number of charged particles in the range $|\eta|<2.4$.
In fig. \ref{fig:cms1}, we compare experimental data \citet{cms}
for pions (black symbols) with the simulations from QGSJETII (upper
left figure), AMPT (upper right), EPOS$\,$LHC (lower left), and EPOS3
(lower right). The different curves in each figure refer to different
centralities, with mean values (from bottom to top) of 8, 84, 160,
and 235 charged tracks. They are shifted relative to each other by
a constant amount. Concerning the models, QGSJETII is the easiest
to discuss, since here there are no flow features at all, and the
curves for the different multiplicities are identical. The data, however,
show a slight centrality dependence: the spectra get somewhat harder
with increasing multiplicity. The other models, AMPT, EPOS$\,$LHC,
and EPOS3 are close to the data. 

In figs. \ref{fig:cms2}, \ref{fig:cms3}, we compare experimental
data \citet{cms} for kaons and protons (black symbols) with the simulations.
The experimental shapes of the $p_{t}$ spectra change considerably,
getting much harder with increasing multiplicity. In QGSJETII, having
no flow, the curves for the different multiplicities are identical.
The AMPT model shows some (but too little) change with multiplicity.
EPOS$\,$LHC goes into the right direction, whereas EPOS3 gives a
reasonable description of the data. \textbf{It seems that hydrodynamical
flow helps considerably to reproduce these data}.

Also ALICE \citet{alice} has measured identified particle production
for different multiplicities in p-Pb scattering at 5.02 TeV. Here,
multiplicity counts the number of charged particles in the range $2.8<\eta_{\mathrm{lab}}<5.1$.
\begin{figure}[tb]
\begin{minipage}[t]{0.45\columnwidth}%
\begin{minipage}[c]{1\columnwidth}%
\vspace*{-0.2cm}

\hspace*{-0.3cm}\includegraphics[angle=270,scale=0.18]{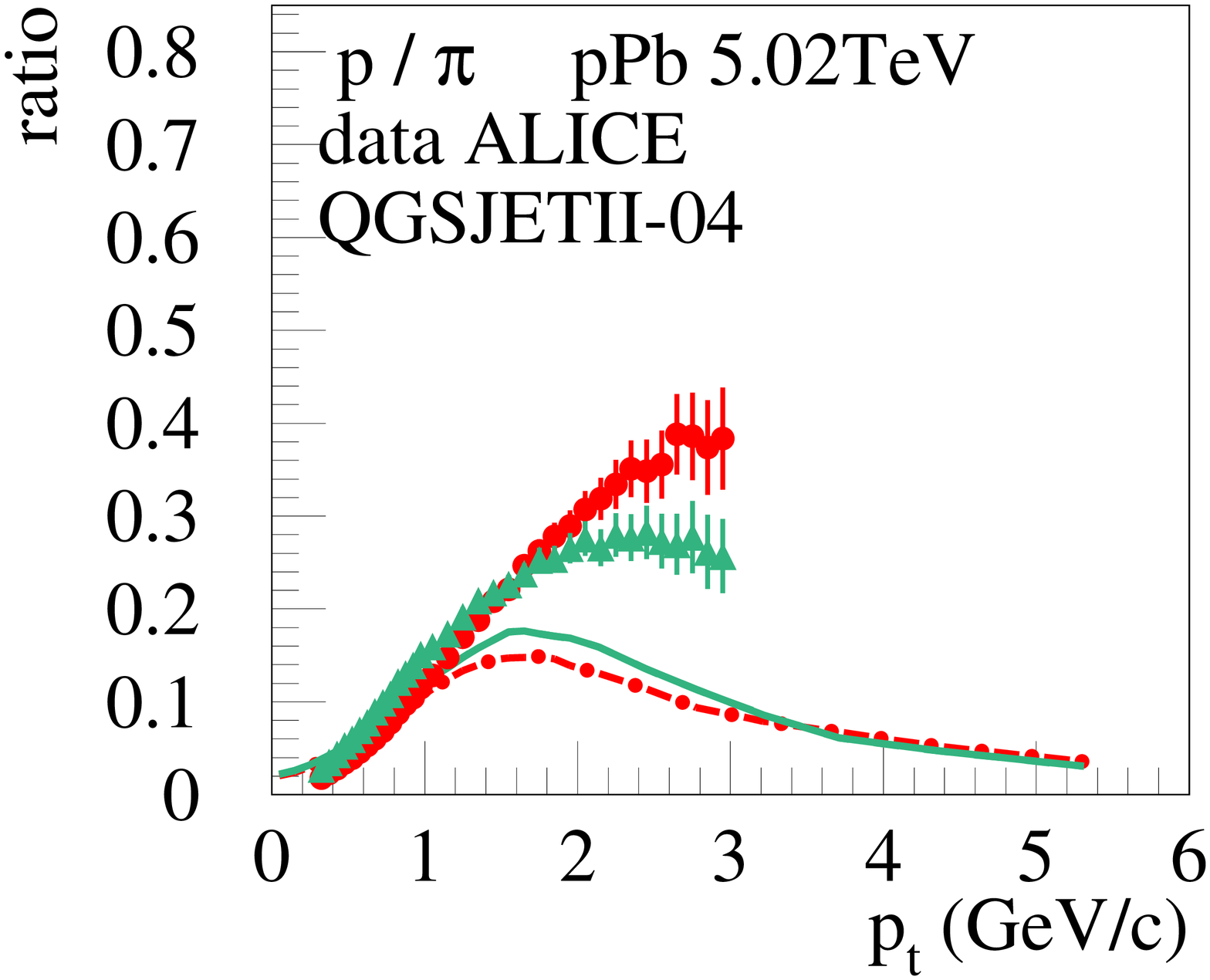}\hspace*{-0.5cm}\includegraphics[angle=270,scale=0.18]{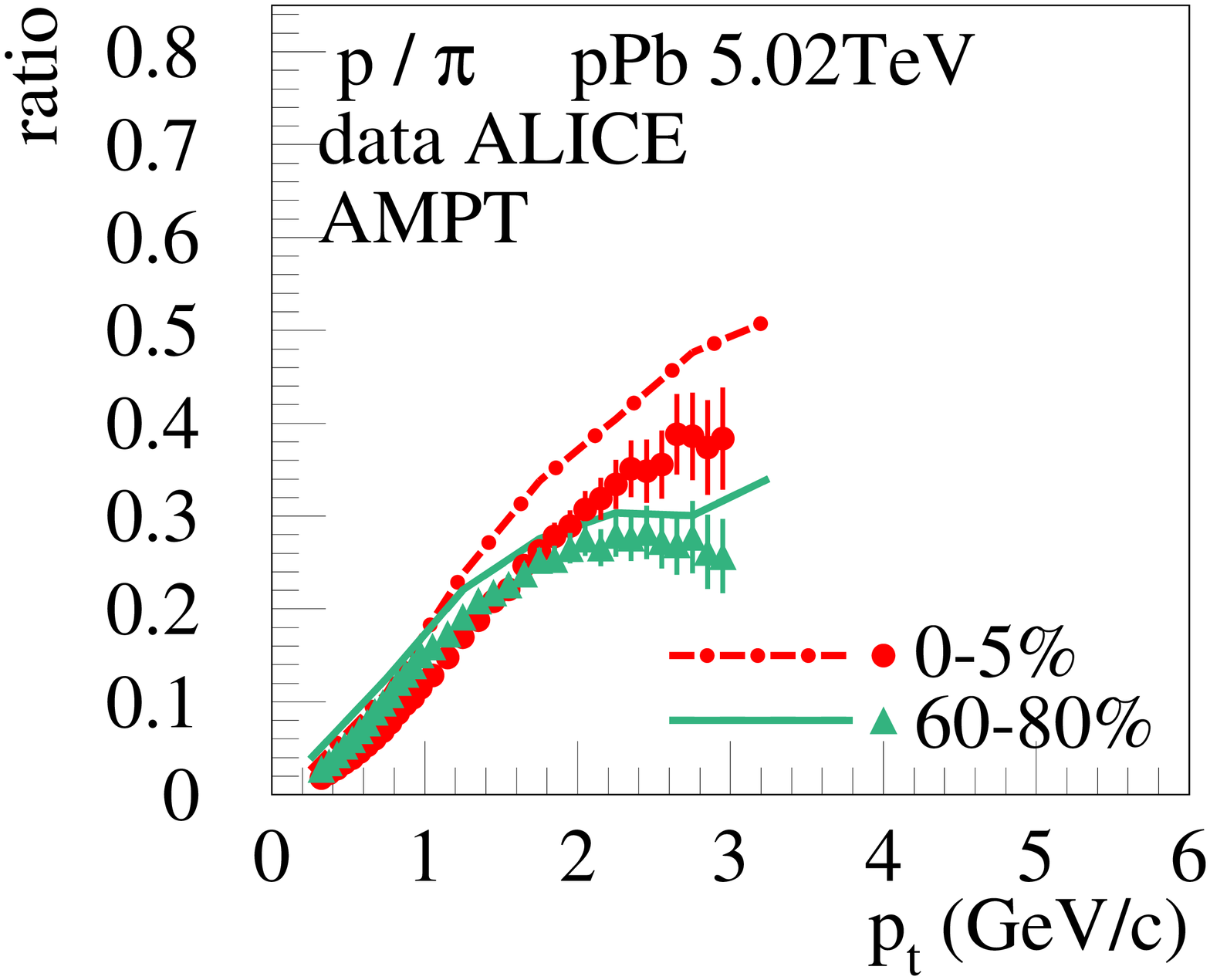}\vspace*{-0.2cm}

\hspace*{-0.3cm}\includegraphics[angle=270,scale=0.18]{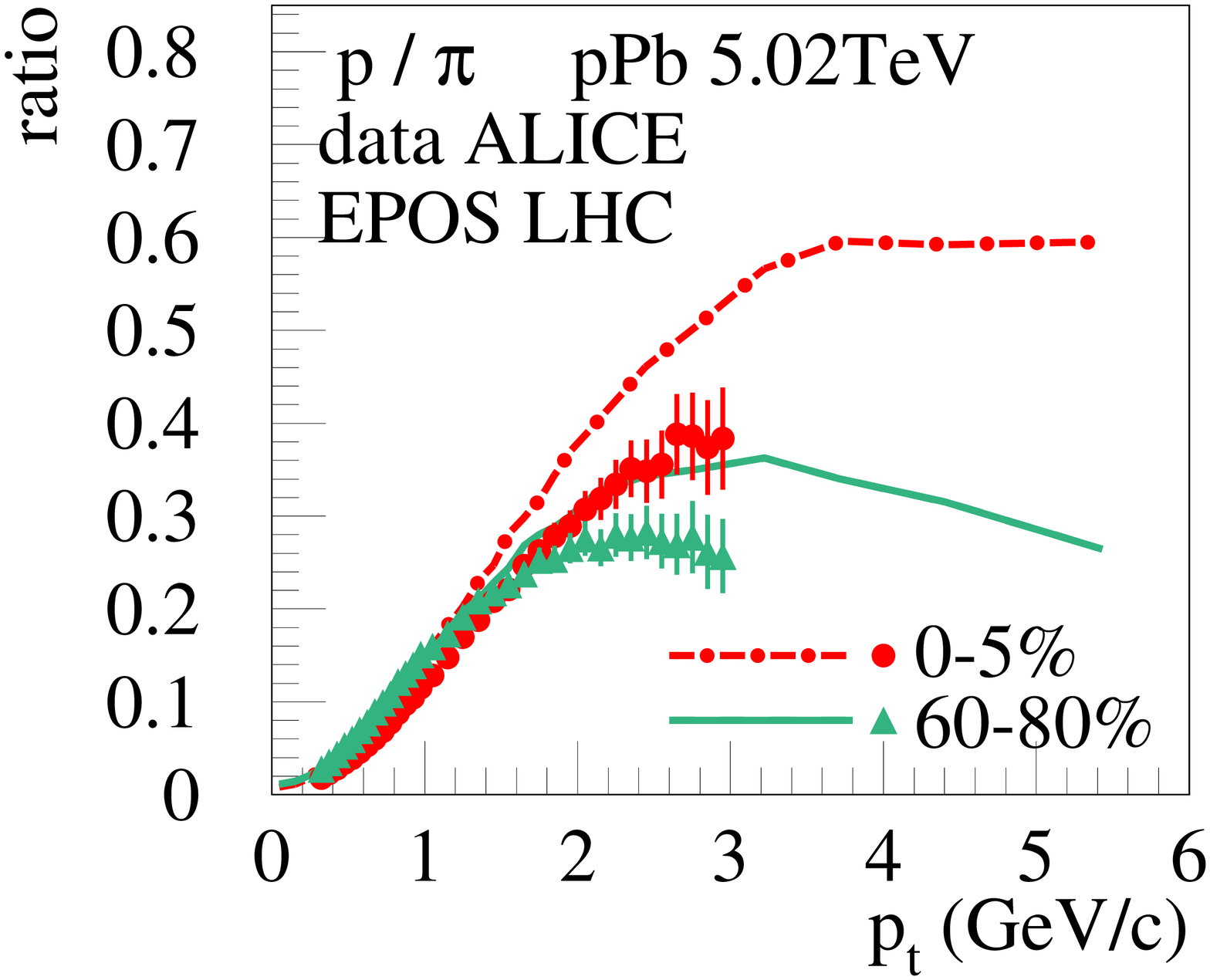}\hspace*{-0.5cm}\includegraphics[angle=270,scale=0.18]{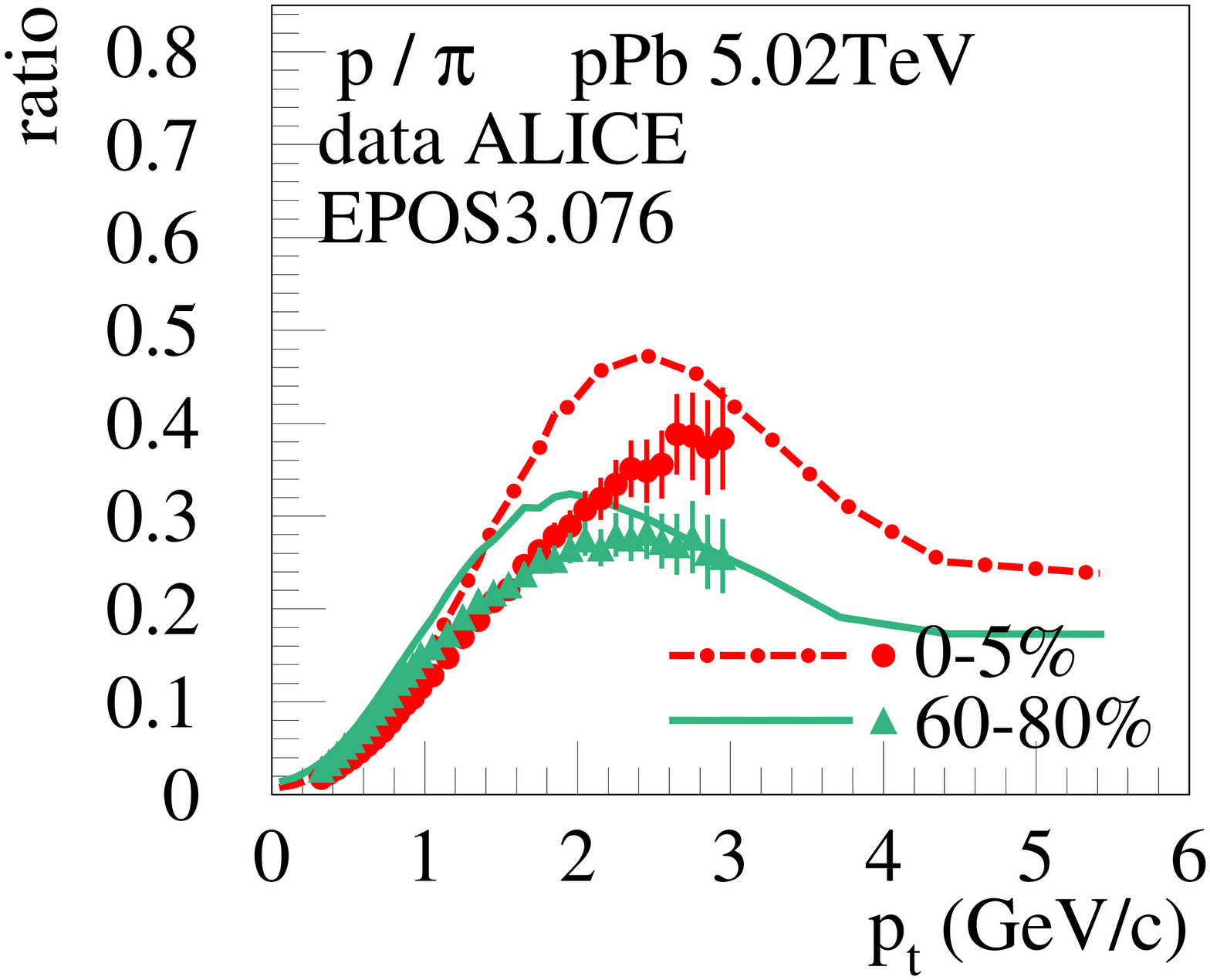}%
\end{minipage}

\noindent \caption{(Color online) Same as fig. \ref{fig:selid07}, but proton over pion
ratio. \label{fig:selid12}}
\end{minipage}$\qquad\quad$%
\begin{minipage}[t]{0.45\columnwidth}%
\begin{minipage}[c]{1\columnwidth}%
\vspace*{-0.2cm}

\hspace*{-0.3cm}\includegraphics[angle=270,scale=0.18]{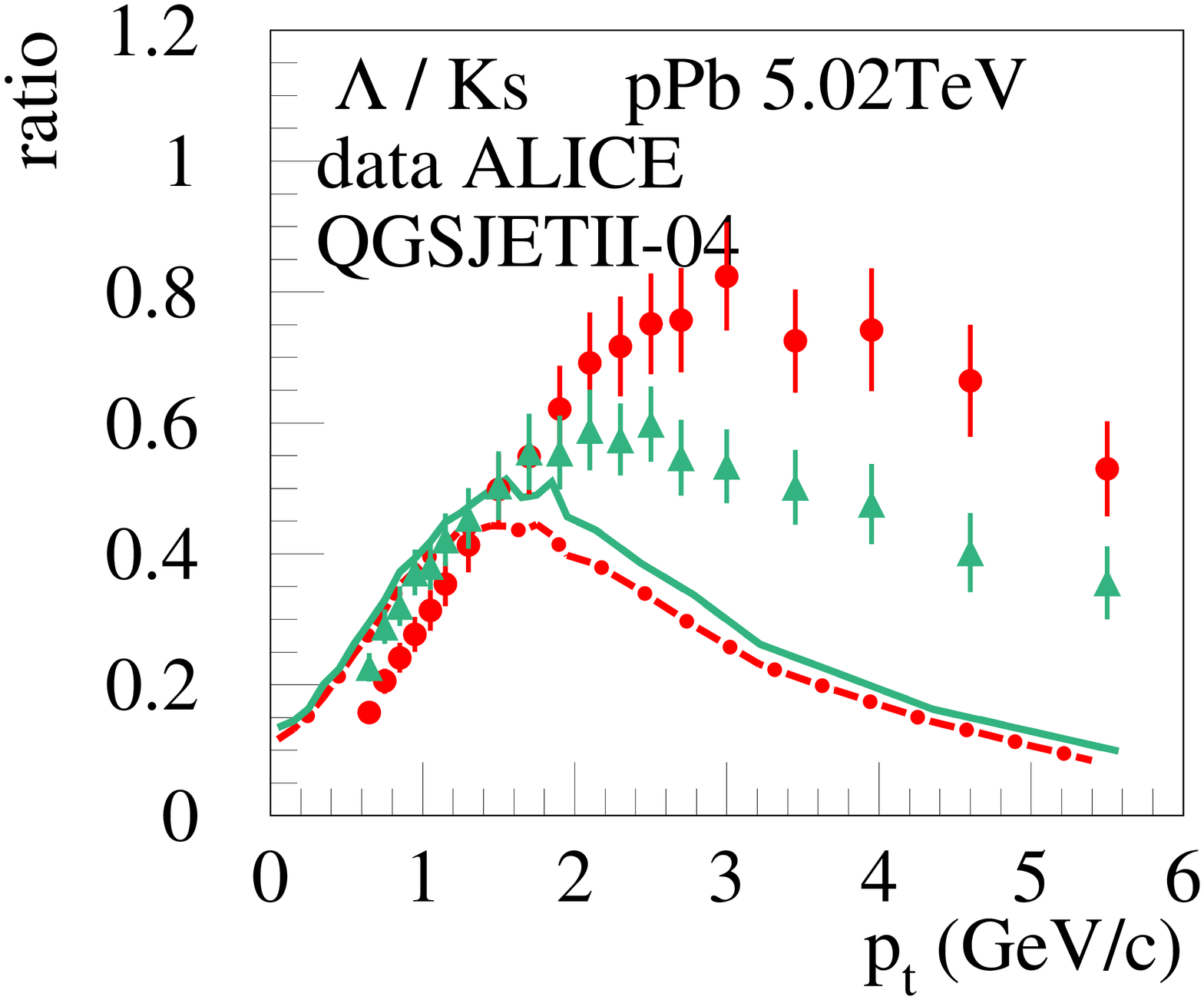}\hspace*{-0.5cm}\includegraphics[angle=270,scale=0.18]{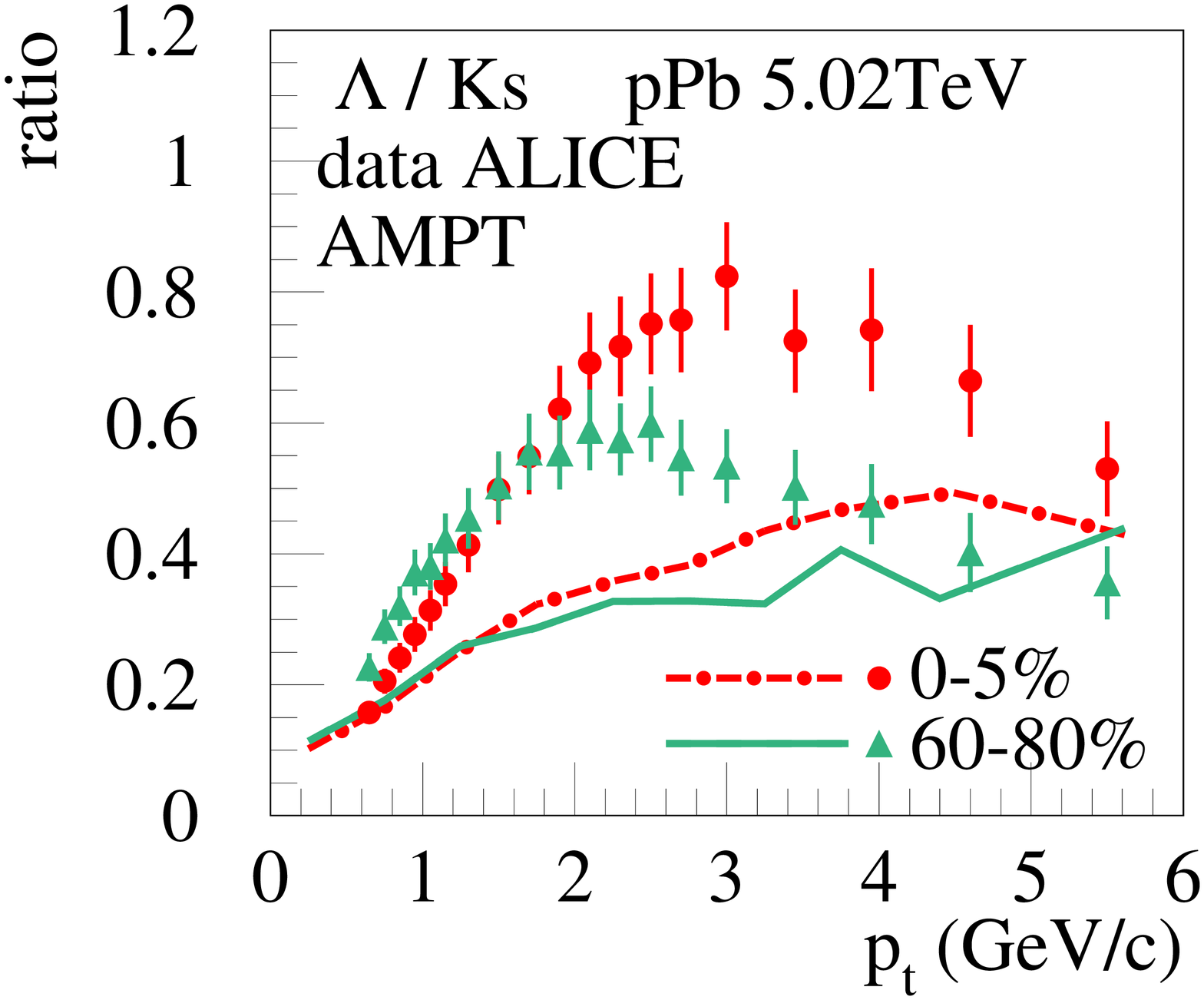}\vspace*{-0.2cm}

\hspace*{-0.3cm}\includegraphics[angle=270,scale=0.18]{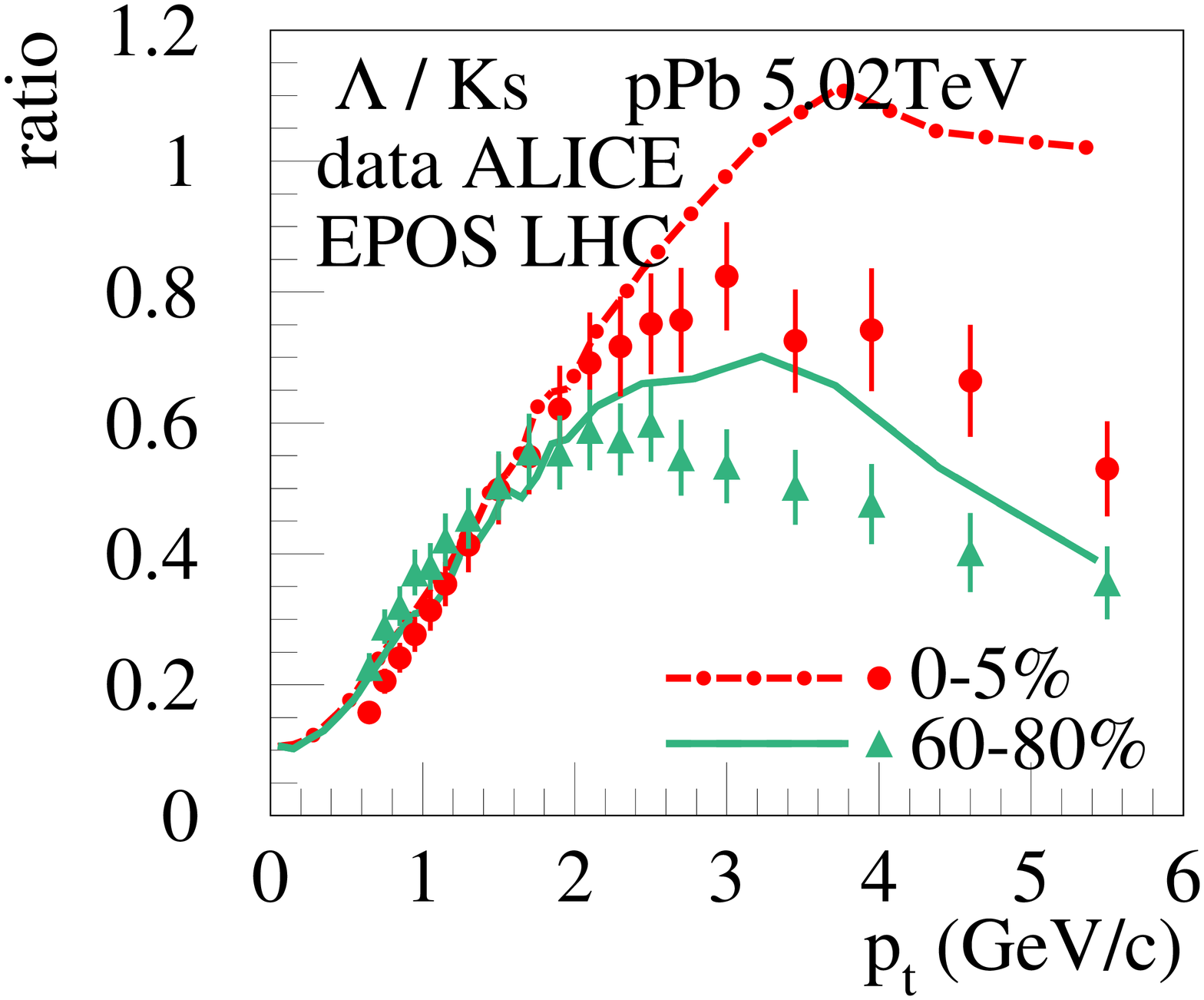}\hspace*{-0.5cm}\includegraphics[angle=270,scale=0.18]{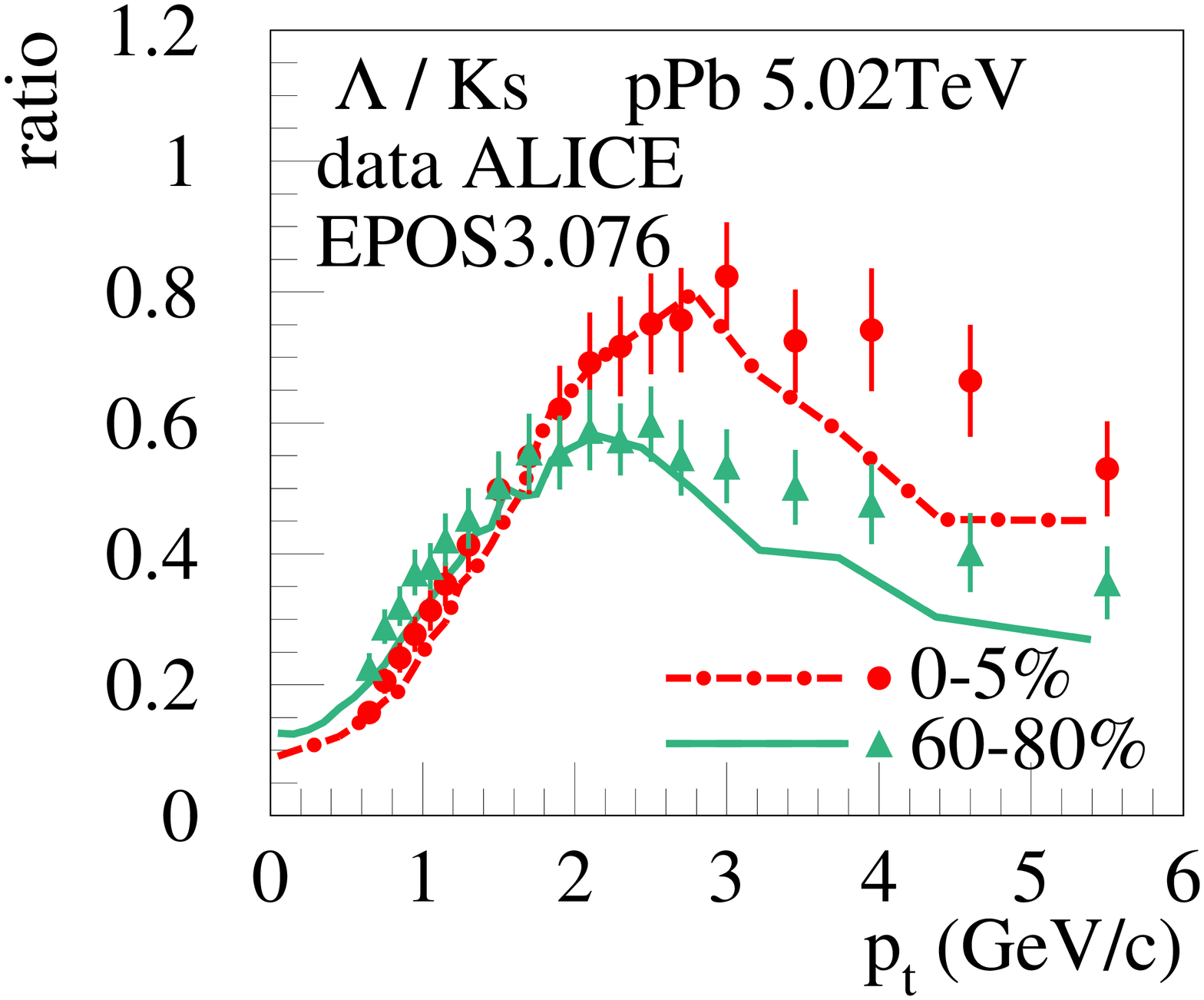}%
\end{minipage}

\noindent \caption{(Color online) Same as fig. \ref{fig:selid07}, but $\Lambda$ over
$K_{s}$ ratio. \label{fig:selid17}}
\end{minipage}%
\end{figure}
It is useful to study the multiplicity dependence, best done by looking
at ratios. In fig. \ref{fig:selid07}, we show the pion over kaon
($K/\pi$) ratio as a function of transverse momentum in p-Pb scattering
at 5.02 TeV, for high multiplicity (red dashed-dotted lines, circles)
and low multiplicity events (green solid lines, triangles), comparing
data from ALICE \citet{alice} (symbols) and simulations from QGSJETII,
AMPT, EPOS$\,$LHC, and EPOS3 (lines). In all models, as in the data,
there is little multiplicity dependence. However, the QGSJETII model
is considerably below the data, for both high and low multiplicity
events. AMPT is slightly below, whereas EPOS$\,$LHC and EPOS3 do
a reasonable job. Concerning the proton over pion ($p/\pi$) ratio,
fig. \ref{fig:selid12}, again QGSJETII is way below the data, for
both high and low multiplicity events, whereas the three other models
show the trend correctly, but being slightly above the data. Most
interesting are the lambdas over kaon ($\Lambda/K_{s}$) ratios, as
shown in fig. \ref{fig:selid17}, because here a wider transverse
momentum range is considered, showing a clear peak structure with
a maximum around 2-3 GeV/c and a slightly more pronounced peak for
the higher multiplicities. QGSJETII and AMPT cannot (even qualitatively)
reproduce this structure. EPOS$\,$LHC shows the right trend, but
the peak is much too high for the high multiplicities. EPOS3 is close
to the data. 

To summarize these ratio plots (keeping in mind that the QGSJETII
model has no flow, AMPT {}``some'' flow, EPOS$\,$LHC a parametrized
flow, and EPOS3 hydrodynamic flow): Flow seems to help considerably.
However, from the $\Lambda/K_{s}$ ratios, we conclude that EPOS$\,$LHC
uses a too strong radial flow for high multiplicity events. The hydrodynamic
flow employed in EPOS3 seems to get the experimental features reasonably
well. Crucial is the core-corona procedure discussed earlier: there
is more core (compared to corona) in more central collisions, but
the centrality (or multiplicity) dependence is not so strong, and
there is already an important core (=flow) contribution in peripheral
events.

\begin{figure}[tb]
\begin{centering}
\includegraphics[angle=270,scale=0.27]{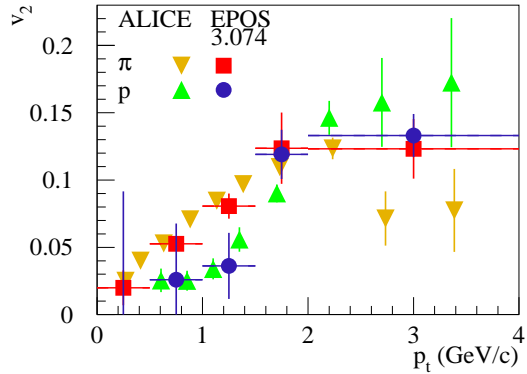}
\par\end{centering}

\caption{(Color online) Elliptical flow coefficients $v_{2}$ for pions and
protons. We show ALICE results (triangles) and EPOS3 simulations (squares
and circles). Pions appear red and yellow, protons blue and green.
\label{fig:v2}}
\end{figure}

Finally, we sketch very briefly results on elliptical flow $v_{2}$
obtained from dihadron correlations, showing ALICE results \citet{alice1,alice2}
and EPOS3 simulation, see ref. \citet{epos3v2} for details. In fig.
\ref{fig:v2}. we plot %
$v_{2}$ as a function of $p_{t}$. Clearly visible in data and in
the simulations: a separation of the results for the tywo hadron species:
in the $p_{t}$ range of 1-1.5 GeV/c, the proton result is clearly
below the pion one. Within our fluid dynamical approach, the above
results are nothing but a {}``mass splitting''. The effect is based
on an asymmetric (mainly elliptical) flow, which translates into the
corresponding azimuthal asymmetry for particle spectra. Since a given
velocity translates into momentum as $m_{A}\gamma v$, with $m_{A}$
being the mass of hadron type $A$, flow effects show up at higher
values of $p_{t}$ for higher mass particles.\\

To summarize : Comparing experimental data on identified particle
production to various Monte Carlo generators, we conclude that hydrodynamical
flow seems to play an important role in p-Pb scattering.\\

This research was carried out within the scope of the GDRE (European
Research Group) {}``Heavy ions at ultrarelativistic energies''.
Iu.K acknowledges support by the National Academy of Sciences of Ukraine
(Agreement 2014) and by the State Fund for Fundamental Researches
of Ukraine (Agreement 2014). Iu.K. acknowledges the financial support
by the ExtreMe Matter Institute EMMI and the LOEWE initiative of the
State of Hesse. B.G. acknowledges the financial support by the TOGETHER
project of the Region of {}``Pays de la Loire''.

~

\end{document}